\documentclass[%
 reprint,
 superscriptaddress,
 groupedaddress,
 showpacs,preprintnumbers,
 bibnotes,
 amsmath,amssymb,
 aps,
 prb,
 floatfix,
]{revtex4-1}

\usepackage{graphicx}
\usepackage{dcolumn}
\usepackage{bm}

\usepackage{color}
\usepackage{hyperref}

\begin{document}
\title{Pressure dependence of the band-gap energy in BiTeI}
\author{S\"{u}meyra G\"{u}ler-K{\i}l{\i}\c{c}}
\email{sumeyra@gtu.edu.tr}
\affiliation{
Department of Physics, Gebze Technical University, Gebze, Kocaeli 41400, Turkey
}
\author{ \c{C}etin K{\i}l{\i}\c{c} }
\email{cetin\_kilic@gtu.edu.tr}
\affiliation{
Department of Physics, Gebze Technical University, Gebze, Kocaeli 41400, Turkey
}


\begin{abstract}
\centerline{\sl Published version available at \url{https://doi.org/10.1103/PhysRevB.94.165203}}
\vspace*{6pt}
The evolution of the electronic structure of BiTeI,
  a layered semiconductor with a van der Waals gap,
  under compression is studied by employing semilocal and dispersion-corrected density-functional calculations.
Comparative analysis of the results of these calculations shows that
  the band-gap energy of BiTeI decreases till it attains a minimum value of {\it zero}
  at a critical pressure, after which it increases again.
The critical pressure corresponding to the closure of the band gap is calculated,
  at which BiTeI becomes a topological insulator.
Comparison of the critical pressure to 
  the pressure at which BiTeI undergoes a structural phase transition
  indicates that the closure of the band gap would not be hindered by a structural transformation.
Moreover, the band-gap pressure coefficients of BiTeI are computed, and
  an expression of the critical pressure is devised in terms of these coefficients.
Our findings
  indicate that
  the semilocal and dispersion-corrected approaches are {\it in conflict} about the compressibility of BiTeI,
  which result in overestimation and underestimation, respectively.
Nevertheless,
  the effect of pressure on the atomic structure of BiTeI is found to be manifested primarily
  as the reduction of the width of the van der Waals gap
  according to both approaches,
  which also yield {\it consistent} predictions concerning the interlayer metallic bonding in BiTeI under compression.
It is consequently shown that
  the calculated band-gap energies follow {\it qualitatively} and {\it quantitatively} the same trend
  within the {\it two} approximations employed here, and
  the transition to the zero-gap state occurs at the {\it same} critical width of the van der Waals gap.
\end{abstract}


\maketitle

Bismuth tellurohalide (BiTeI)
  is a narrow-band-gap semiconductor
  at ambient pressure.
Under compression,
  the BiTeI band gap decreases with increasing pressure, and
  diminishes at a critical pressure $P_c$.
Recent studies show that
  an inversion of valence band maximum (VBM) and conduction band minimum (CBM) states 
  occurs when the applied pressure $P$ exceeds $P_c$,
  which marks a transition to the topological insulating phase.\cite{bahramy12, xi13, ideue14, park15}
This means that
  the band gap reopens and increases at pressures higher than $P_c$.
Furthermore, analysis presented in Ref.~\onlinecite{liu14} indicates that                       
  an intermediate Weyl semimetal phase is present over a narrow pressure interval around $P_c$. 
The evolution of the electronic structure of BiTeI under compression                            
  has been experimentally explored in two optical spectroscopy studies:\cite{xi13,tran14}       
Tran {\it et al.} (Ref.~\onlinecite{tran14}) reported {\it no} evidence for the band-gap reopening,
  whereas the infrared spectral weight of the charge carriers measured in Ref.~\onlinecite{xi13} exhibited a maximum,
  which implies the reopening of the band gap with increased pressure. 
On the other hand,
  both studies\cite{xi13,tran14} reveal that
  BiTeI undergoes a structural transformation at pressure $P_t \sim$8 GPa (Ref.~\onlinecite{xi13}) or
                                                              $\sim$9 GPa (Ref.~\onlinecite{tran14}).
It was therefore suggested that the topological phase transition
  is hindered by a structural phase transition.\cite{tran14}
On the theoretical side,
  density-functional calculations have been employed to estimate $P_c$,
  the reported values of which range from 1.7 to $\sim$10 GPa
  owing to various types of approximations.\cite{bahramy12,chen13,tran14,ideue14,park15}
It is crucial to have a reliable estimate of $P_c$ to see
  if the topological phase transition takes place before the structural phase transition occurs,
  which clearly depends critically on the accuracy of the equation of state (EOS).
Our recent investigation\cite{kilic15} revealed that
  the inclusion of van der Waals (vdW) interactions is necessary for a reliable and truly {\it ab initio} computation of the EOS of BiTeI,
  which was not taken into account in the aforementioned calculations.
Thus, we examine here
  the variation of the BiTeI band gap with pressure
  with the aid of dispersion-corrected density-functional (PBE-D2) calculations
  using the functional of Perdew, Burke, and Ernzerhof (PBE)\cite{perdew96}
  together with a semiempirical force field.\cite{grimme06}
In addition, we employ semilocal density-functional calculations
  using the PBEsol\cite{perdew08} functional for the purpose of comparison.
It was curious to see whether
  the PBEsol and PBE-D2 approaches
  result in {\it consistent} predictions for BiTeI,
  which yield qualitatively similar results for systems with vdW-bonded layers.\cite{tunega12}
Our findings
  indicate that
  the PBEsol and PBE-D2 approaches are {\it in conflict} about the compressibility of BiTeI,
  which nevertheless yield {\it consistent} predictions concerning the evolution of the electronic structure of BiTeI under compression,
  as will be revealed below.

BiTeI crystallizes in a layered trigonal structure
  with noncentrosymmetric space group $P3m1$ (No. 156).
Its crystal structure is characterized by the hexagonal lattice parameters
  $a$ and $c$, and two internal parameters
  since Bi, Te, and I atoms occupy the $1a$, $1c$, and $1b$ positions
  with fractional coordinates (0,0,0),
                              (2/3,1/3,$z_{\rm Te}$), and
                              (1/3,2/3,$z_{\rm I}$), respectively,\cite{shevelkov95}
  cf. Ref.~\onlinecite{kilic15}.
In this structure, adjacent (unary) layers formed by Bi, Te, and I atoms stack along the $c$-axis,
  and the {\it van der Waals (vdW) gap} exists in the vacuum region between the Te and I layers.
As will be clear below,
  our findings reveal the importance of the width of the vdW gap.
Thus we investigate the effect of pressure not only on the BiTeI band structure
  but also on the vdW gap.
To this end,
  we carried out crystal-structure optimizations,
  for which the total energy was computed as function of the unit cell volume $\Omega$ and
  the lattice parameter ratio $c/a$,
  followed by band-structure calculations.
For each value of the pair ($\Omega$, $c/a$),
  the ionic positions were relaxed until
  the maximum value of residual forces on atoms was reduced to be smaller than 0.01 eV/\AA.
The total-energy and band-structure calculations were performed
  by employing the projector augmented-wave (PAW) method,\cite{blochl94} as implemented\cite{kresse99}
  in the Vienna {\it ab initio} simulation package (VASP).\cite{kresse96}
Spin-orbit coupling (SOC) was taken into account
  by utilizing the noncollinear mode of VASP.\cite{hobbs00,marsman02}
The 5$s$ and 5$p$,
    5$s$ and 5$p$, and
    6$s$ and 6$p$
    states are treated as valence states for
    tellurium,
    iodine, and
    bismuth, respectively.
Plane wave basis sets were used to represent the electronic states, which were determined by
  imposing a kinetic energy cutoff of 325 eV.
The Brillouin zone was sampled by
  $20\times 20\times 16$ {\bf k}-point mesh in total-energy calculations,
  which was generated according to the Monkhorst-Pack scheme.\cite{monkhorst76}
Convergence criterion for the electronic self-consistency was set up to
  10$^{-6}$~eV (10$^{-8}$~eV) in total-energy (band-structure) calculations.
The parameters used for the semiempirical (van der Waals) force field\cite{grimme06}
  were the same as in Ref.~\onlinecite{kilic15}.
It should be mentioned that although we were inclined to employ a nonempirical van der Waals density functional 
  (in lieu of the semiempirical force field)                                                               
  such as the optB86b-vdW functional,\cite{klimes11}                                                       
  we unfortunately found that                                                                              
  the crystal structure of BiTeI was not adequately described by using the optB86b-vdW functional,         
  as discussed in the Appendix.                                                                            

\begin{figure}
  \begin{center}
    \resizebox{0.48\textwidth}{!}{
      \includegraphics{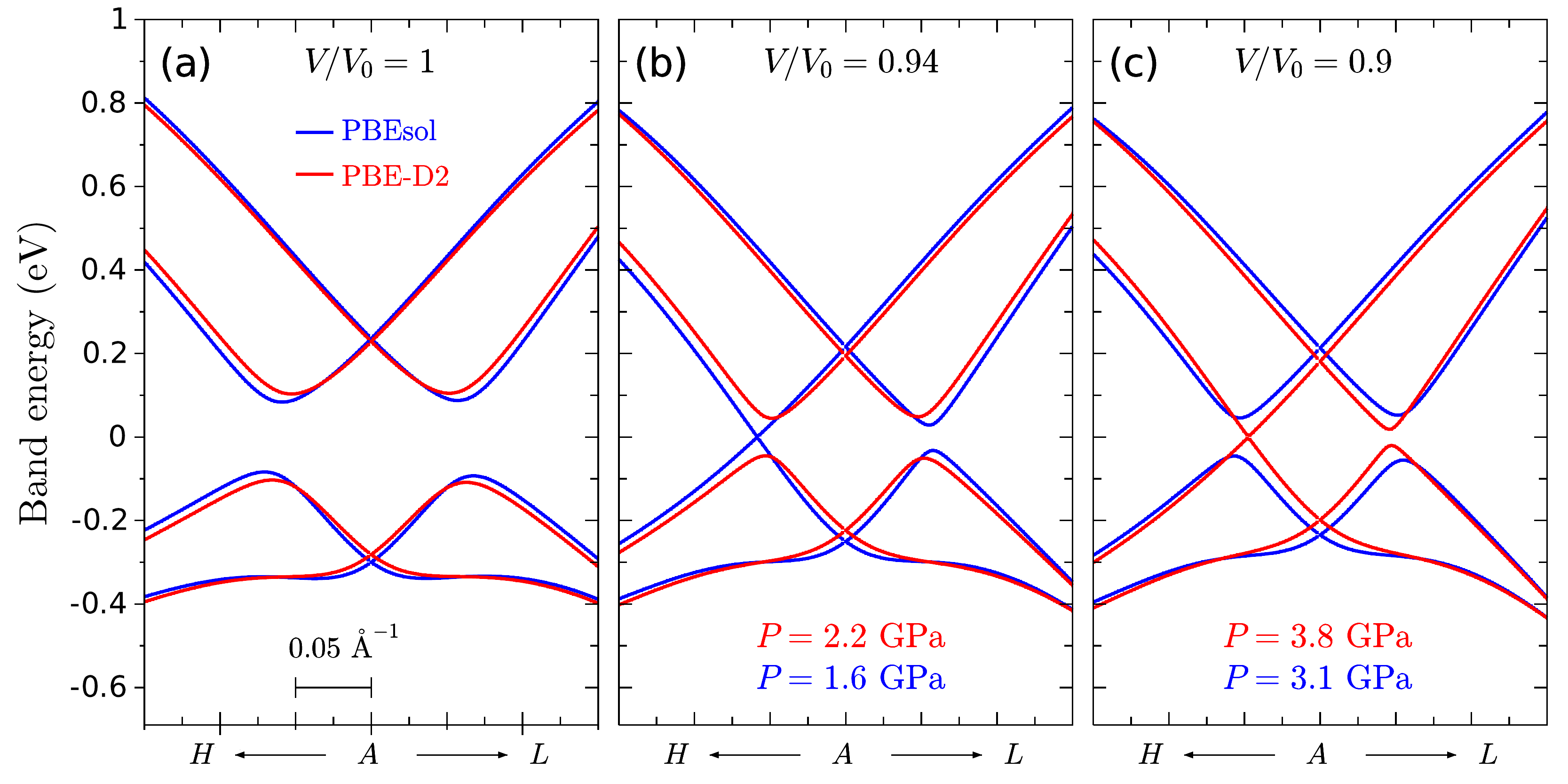}
    }
  \end{center}
  \vspace*{-0.5cm}
  \caption{ The band structure of BiTeI in the vicinity of the band edges for (a) $V=V_0$,
                                                                                            (b) $V=0.94 V_0$, and
                                                                                            (c) $V=0.9  V_0$.}
  \vspace*{-0.5cm}
  \label{f:bs}
\end{figure}

\begin{figure}
  \begin{center}
    \resizebox{0.5\textwidth}{!}{
      \includegraphics{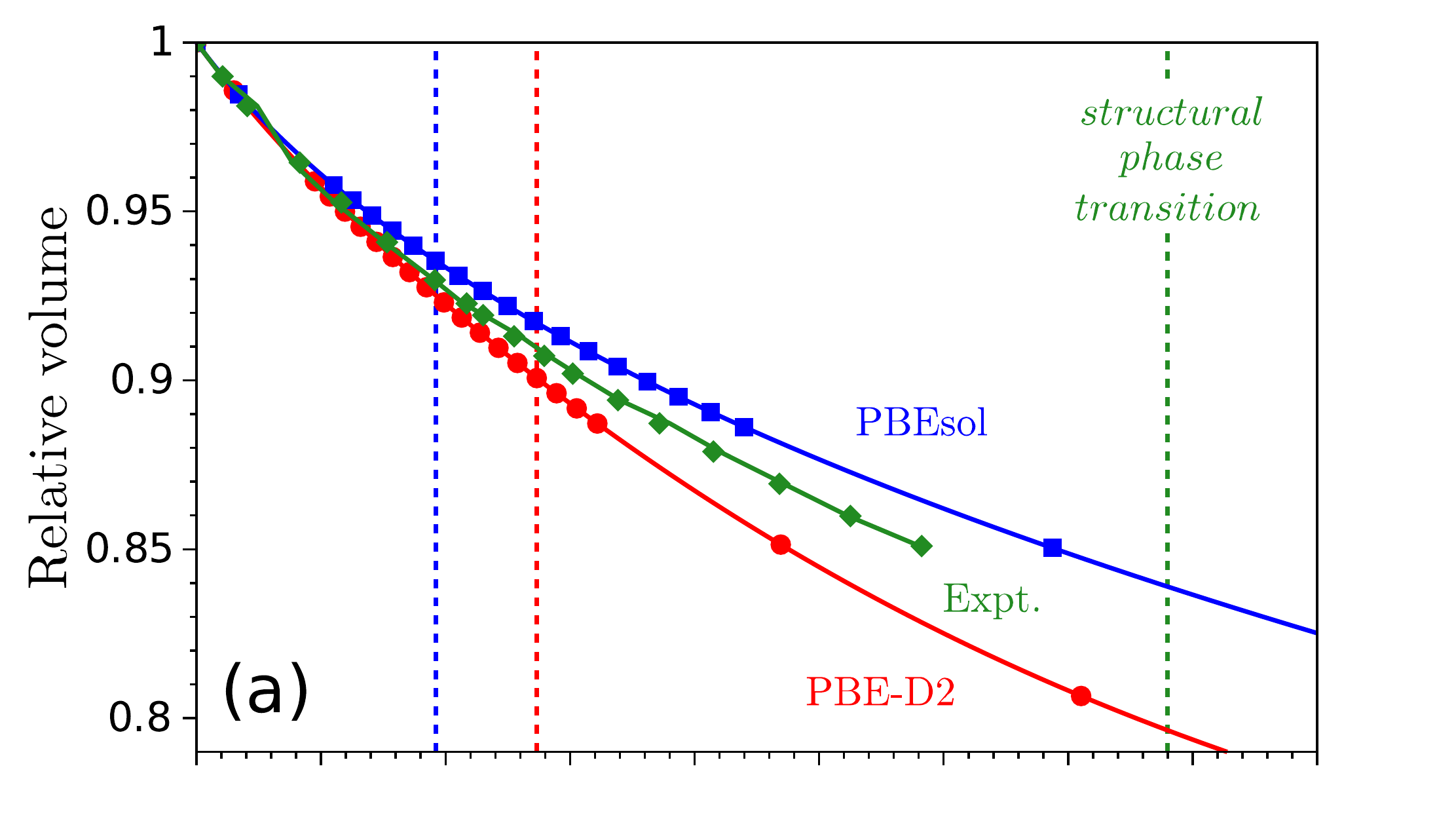}
    }
    \resizebox{0.5\textwidth}{!}{
      \includegraphics{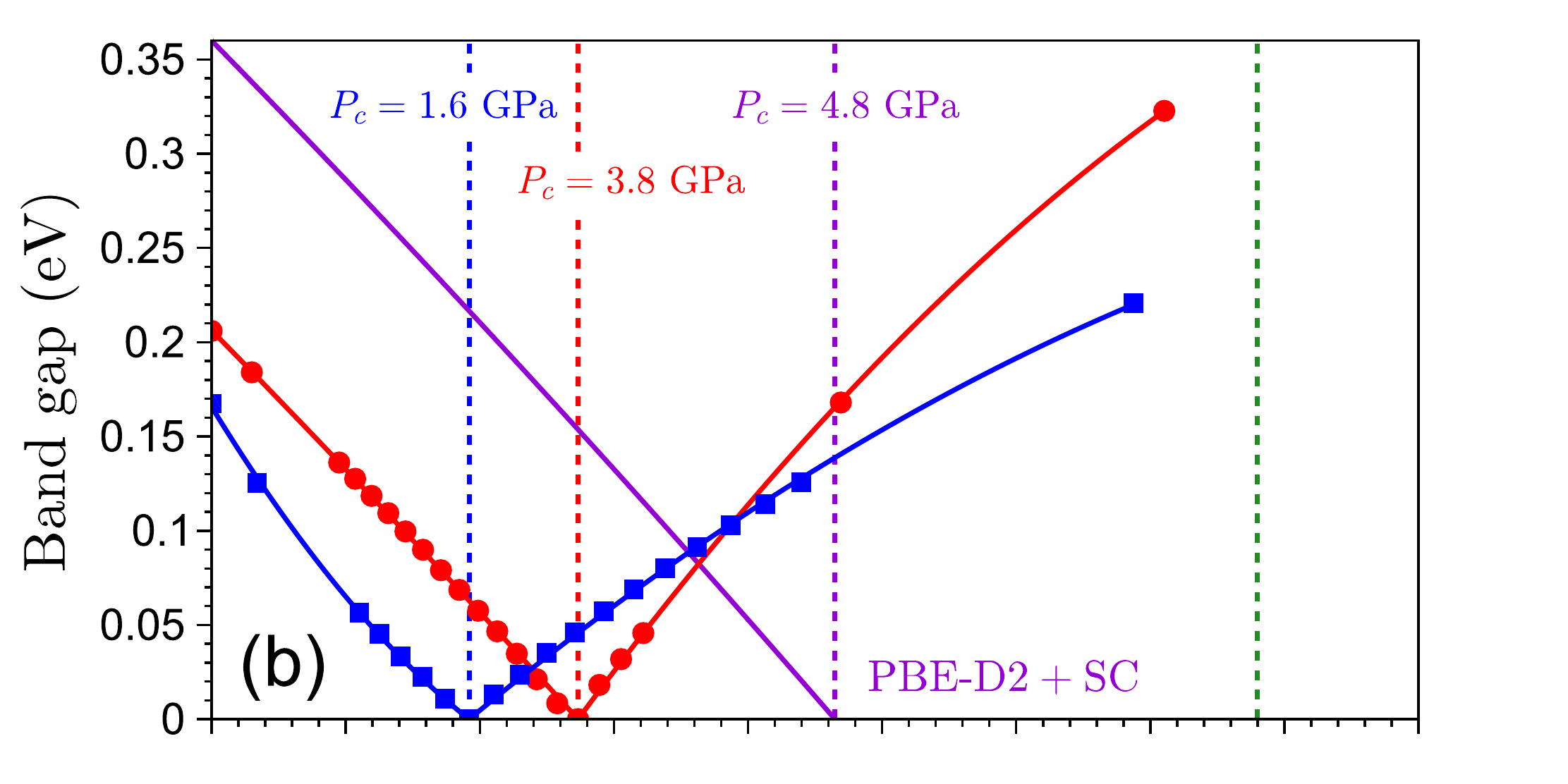}
    }
    \resizebox{0.5\textwidth}{!}{
      \includegraphics{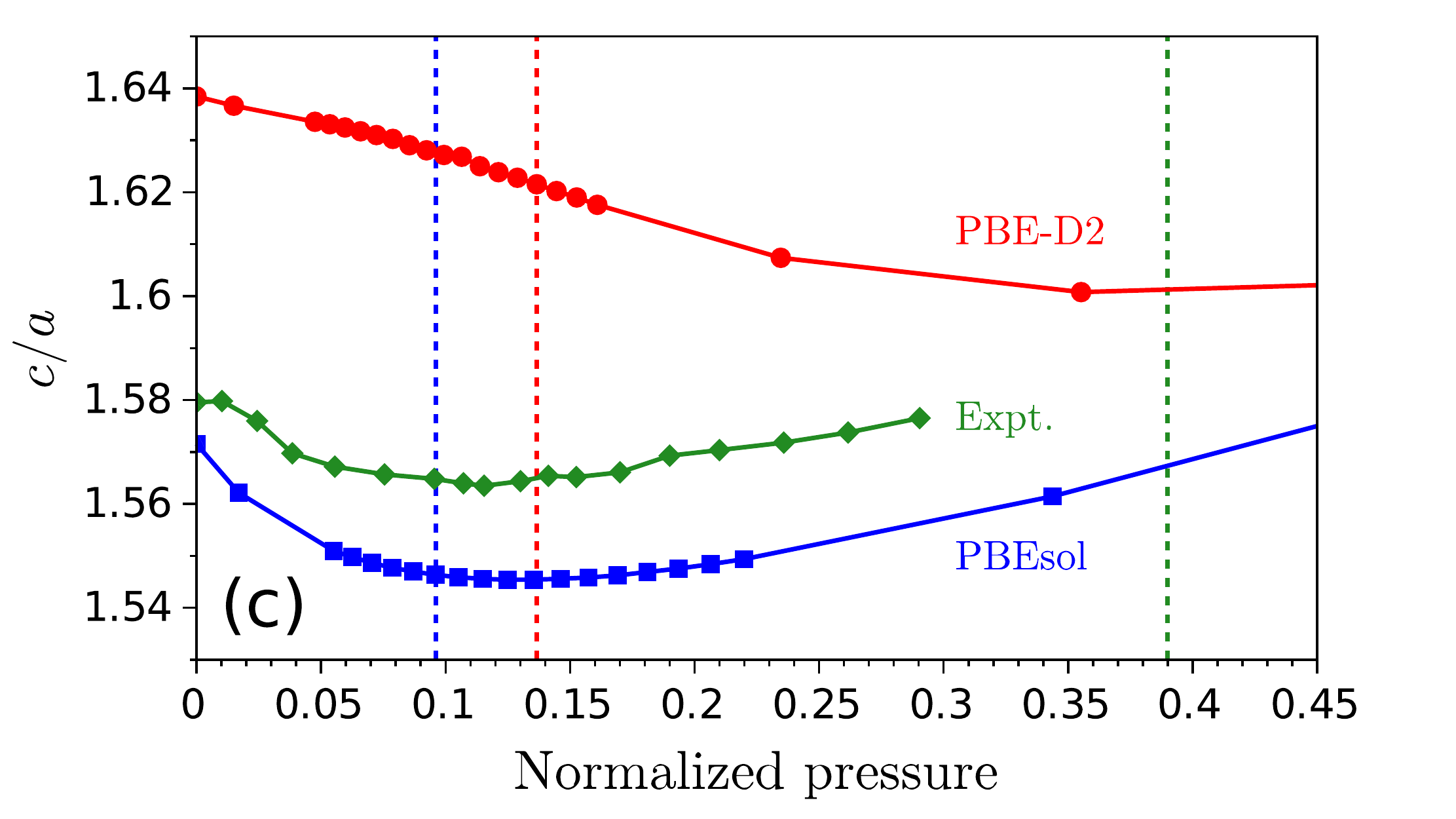}
    }
  \end{center}
  \vspace*{-0.5cm}
  \caption{ The plots of the relative volume $V/V_0$ (a),
                                               the band-gap energy $E_g$ (b) and
                                the lattice parameter ratio $c/a$ (c)
           versus the normalized pressure $P/K_0$.
           In all panels, the PBEsol-calculated, experimental\cite{xi13,kilic15} and dispersion-corrected (PBE-D2) values
             are given in blue, green and red, respectively.
           The vertical blue and red dashed lines mark the calculated (PBEsol and PBE-D2) values of $P_c/K_0$.
           The vertical green dashed line marks the experimental value of $P_t/K_0$.
           In (b), the ``scissors''-corrected (PBE-D2+SC) results are given in magenta.
          }
  \vspace*{-0.25cm}
  \label{f:eos}
\end{figure}

Representative results of band-structure calculations are given in Figs.~\ref{f:bs}(a)-\ref{f:bs}(c)
  where $V_0$ denotes the equilibrium value of the volume $V$ per formula unit,
  and $P$ is the pressure obtained from the calculated EOS (see below).
It is known that
  SOC-induced modification of the BiTeI band edges
  leads to large Rashba-type splitting\cite{ishizaka11} near the Brillouin zone point $A$,
  which is visible Figs.~\ref{f:bs}(a).
Furthermore, strong (weak) SOC leads to a smaller (larger) band gap 
  as well as a greater (smaller) Rashba coupling strength.\cite{kilic15}
The decreasing behavior of the band gap, cf. Figs.~\ref{f:bs}(a)-\ref{f:bs}(c), 
  could therefore be
  partly                                                                      
  attributed to increased SOC at higher compression,
  which also reflects the increase in the valence- and conduction-band width. 
As seen in Figs.~\ref{f:bs}(b) and \ref{f:bs}(c),
  the transition to the semimetallic state
  occurs at a considerably lower pressure $P_c=1.6$~GPa [Fig.~\ref{f:bs}(b)] within the PBEsol approach,
                              compared to the respective dispersion-corrected (PBE-D2) value of $P_c=3.8$~GPa [Fig.~\ref{f:bs}(c)].
It is thus hard to claim that the predicted (PBEsol and PBE-D2) values of the critical pressure
  agree with each other.
To expound on the origin of this disagreement,
  we compare the calculated EOS curves to each other and to the experimental EOS.
Since the calculated and experimental EOSs refer to zero and room temperature, respectively,
  it is appropriate\cite{kilic15} to use the relative volume $V/V_0$  and the normalized pressure $P/K_0$,
  with $K_0$ denoting the bulk modulus.
Thus a comparison of the $V/V_0$ versus $P/K_0$ curves is given in Fig.~\ref{f:eos}(a),
  where the experimental curve is clearly {\it bracketed} by the PBEsol-calculated and dispersion-corrected (PBE-D2) curves.
The latter were obtained by performing
  (i)   third-,
  (ii)  fourth-, and
  (iii) fifth-order Birch-Murnaghan (BM) fits to
  (i)   the experimental compressibility data,\cite{xi13}
  the energy-volume curve obtained via
  (ii)  PBEsol, and
 (iii)  PBE-D2 calculations, respectively.
The higher-order BM fits are necessary\cite{kilic15} to preserve the level of accuracy
  of the PBEsol or PBE-D2 calculations, cf. Fig.~S1 (Ref.~\onlinecite{supmat}).
Thus the pressure $P$ is given by
  \begin{equation}
    P = 3K_0f(1+2f)^{5/2} \left ( 1 + a_1 f + a_2 f^2 + a_3 f^3 \right )
  \end{equation}
where  $a_1=(3/2)(K^\prime_0-4)$,
       $a_2=(3/2)( K_0K^{\prime\prime}_0 + K^\prime_0(K^\prime_0-7) + 143/9 )$,
       $a_3=(1/8)[9K_0^2 K^{\prime\prime\prime}_0 + 12K_0(3K^\prime_0-8)K^{\prime\prime}_0 + K^\prime_0[(3K^\prime_0-16)^2+118]-1888/3 ]$,
and $f= [ ( V_0/V  )^{2/3} - 1 ]/2$ is the Eulerian strain.
Here $K^\prime_0$, $K^{\prime\prime}_0$, and $K^{\prime\prime\prime}_0$ denote
  the first, second, and third pressure derivatives of the bulk modulus, respectively.
The predicted values of $V_0$ are 111.716 \AA$^3$ (PBEsol) and
                                  111.585 \AA$^3$ (PBE-D2),
  which are in close agreement with each other as well as with the experimental value\cite{shevelkov95} of 111.762 \AA$^3$.
The calculated and experimental values of the bulk modulus and its pressure derivatives are given in Table~\ref{t:K0}.
Note that the experimental values of $K_0$ and $K_0^\prime$
  are {\it bracketed} by the respective PBEsol-calculated and dispersion-corrected (PBE-D2) values.
It is also notable that
  BiTeI is significantly more compressible within the PBEsol approach,
    compared to the dispersion-corrected (PBE-D2) approach,
  since $K_0^{\rm PBEsol}/K_0^{\rm PBE\mbox{-}D2}=0.6$.
As discussed further below,
  this explains why $P_c^{\rm PBEsol}$ is significantly smaller than $P_c^{\rm PBE\mbox{-}D2}$.

\begin{table}
\caption{\label{t:K0}
        The calculated (PBEsol and PBE-D2) and experimental\cite{xi13,kilic15} values of
        the bulk modulus $K_0$ (in GPa) and its pressure derivatives 
                         $K_0^{\prime}$, 
                         $K_0^{\prime\prime}$ (in GPa$^{-1}$), and 
                         $K_0^{\prime\prime\prime}$ (in GPa$^{-2}$)
        of BiTeI.
        }
\begin{ruledtabular}
\begin{tabular}{lcccc}
            & $K_0$ & $K_0^{\prime}$ & $K_0^{\prime\prime}$ & $K_0^{\prime\prime\prime}$ \\ \hline
  PBE-D2    & 28.1  &  6.8           &  -2.9                &   2.5                      \\
  PBEsol    & 17.0  & 12.4           &  -7.2                &                            \\
  Expt.     & 20.5  &  7.6           &                      &                            
\end{tabular}
\end{ruledtabular}
\end{table}

As explained in Ref.~\onlinecite{liu14},                                                
  the BiTeI band gap $E_g$ is {\it zero} for $P_c \leqslant P \leqslant P_c+\Delta P_c$ 
  with $\Delta P_c > 0$,                                                                
  where $\Delta P_c$ denotes                                                            
  the width of the pressure interval corresponding to the Weyl semimetal phase.         
Analysis presented in Ref.~\onlinecite{liu14}                                           
 reveals also that one finds $\Delta P_c = 0$                                           
  {\it if} $E_g$ is deduced {\it only} from the band dispersions                        
  along the $A$-$H$ line in the irreducible wedge of the Brillouin zone.                
Thus we conducted a search                                                              
  over the {\it whole} Brillouin zone                                                   
  for a set of pressure values around the critical pressure.                            
Although we had not attempted to deduce a precise value for $\Delta P_c$,               
  in this way we estimated upper and lower bounds for it:                               
  0.04~GPa $< \Delta P_c <$ 0.13 GPa (PBE-D2) and                                       
  0.03~GPa $< \Delta P_c <$ 0.09 GPa (PBEsol).                                          
Clearly,                                                                                
  the experimental observation of                                                       
  the Weyl semimetal phase\cite{liu14}                                                  
  as well as the closure of the band gap                                                
  would be obstructed, cf. Ref.~\onlinecite{tran14},                                    
  unless the applied pressure is fine tuned to this extremely narrow pressure range.    
We shall defer                                                                          
  the discussion of the Weyl semimetal phase in the rest of this paper,                 
  due to the diminutiveness of $\Delta P_c$.                                            

Figure~\ref{f:eos}(b) displays a plot of the band-gap energy $E_g$ versus the normalized pressure $P/K_0$,
  which shows that the band gap diminishes at $P=P_c$ and reopens for $P>P_c$.
The predicted band gap takes the values
  0.167 eV (PBEsol) and 0.206 eV (PBE-D2) at zero pressure,
  compared to the experimental\cite{lee11} value of 0.36 eV at ambient pressure.
This shows that the band-gap underestimation
  is relatively more (less) severe within the PBEsol (PBE-D2) description.
The solid symbols (blue squares and red circles) in Fig.~\ref{f:eos}(b)
  represent the calculated (PBEsol and PBE-D2) points.
The blue and red curves passing through these points
  satisfy the relationship
\begin{equation}\label{eq:EgP}
  E_g(P)=C_1\left (\frac{P}{K_0}-\frac{P_c}{K_0}\right )+C_2\left (\frac{P}{K_0}-\frac{P_c}{K_0}\right )^2,
\end{equation}
where the coefficients $C_1$ and $C_2$ take different values for the cases $P<P_c$ and $P_c<P$.
The values of these coefficients are given in Table~\ref{t:Pc},
  which were obtained via fitting.
It is noteworthy to observe that 
  the $P<P_c$ portions of blue (PBEsol) and red (PBE-D2) curves in Fig.~\ref{f:eos}(b) are almost parallel to each other,
  especially in the vicinity of the critical pressure $P_c$.
This implies that this portion of the dispersion-corrected (PBE-D2) curve could be qualitatively reproduced
  by adding a constant shift to the PBEsol-calculated band-gap energies,
  which is known as the ``scissors'' correction.
Encouraged by the latter,
  we apply the ``scissors'' correction
  to the dispersion-corrected (PBE-D2) band-gap energies
  by using a constant shift of 154 meV
  that is equal to the difference between the experimental\cite{lee11} and predicted values of the band gap.
In Fig.~\ref{f:eos}(b)
  the magenta curve represents
  the ``scissors''-corrected (PBE-D2+SC) band-gap energy as a function of the normalized pressure.
It is seen that
  the predicted value of the critical pressure increases from $P_c=3.8$ to $P_c=4.8$~GPa
  as a result of applying the ``scissors'' correction.
Thus, underestimation of the band gap by 154 meV
  translates to underestimating $P_c$ by 1 GPa.
It should be emphasized that
  {\it all} of our estimates for $P_c$ are substantially smaller than $P_t$, cf. Fig.~\ref{f:eos}(b).
It is thus clear that the closure of the band gap would not be hindered by the structural transformation of BiTeI at $P=P_t$.

\begin{table}
\caption{\label{t:Pc}
         The calculated (PBEsol and PBE-D2) values of 
         the equilibrium band gap $E_g$ (in meV),
         the critical pressure $P_c$ (in GPa), and
         the coefficients $C_1$ and $C_2$ introduced in Eq.~(\ref{eq:EgP}).
        }
\begin{ruledtabular}
\begin{tabular}{lccccr}
       & $E_g$ & $P_c$ & $C_1$ & $C_2$ &             \\ \hline
PBE-D2 & 206   & 3.8   &            &            &             \\
       &       &       &  -1.565    & -0.407     & for $P<P_c$ \\
       &       &       & \ 1.926    & -2.057     &     $P>P_c$ \\
PBEsol & 167   & 1.6   &            &            &             \\
       &       &       &  -1.127    &\ 6.245     & for $P<P_c$ \\
       &       &       & \ 1.176    & -1.160     &     $P>P_c$
\end{tabular}
\end{ruledtabular}
\end{table}

\begin{table}
\caption{\label{tab:alpha}
         The calculated (PBEsol and PBE-D2) values of
         the band-gap pressure coefficients $\alpha_1$ (in meV/GPa) and $\alpha_2$ (in meV/GPa$^2$).
        }
\begin{ruledtabular}
\begin{tabular}{lrr}
       & $\alpha_1$ & $\alpha_2$    \\ \hline
PBE-D2 &  -51.8     &  -1.0         \\
PBEsol & -135.4     &  43.2         \\
\end{tabular}
\end{ruledtabular}
\end{table}

It should be noted that the band-gap pressure coefficients $\alpha_1=dE_g/dP|_{P=0}$
                                                       and $\alpha_2=d^2E_g/dP^2|_{P=0}$
are related to the coefficients introduced in Eq.~(\ref{eq:EgP})
through $\alpha_1=C_1/K_0-2C_2P_c/K_0^2$
    and $\alpha_2=C_2/K_0^2$.
The calculated  values of $\alpha_1$ and $\alpha_2$ are given in Table~\ref{tab:alpha}.
Rephrasing Eq.~(\ref{eq:EgP}) as $E_g(P)=E_g+\alpha_1 P+(1/2)\alpha_2 P^2$ (for $P \le P_c$)
  yields the following expression for the critical pressure:
\begin{equation}\label{eq:Pc}
  P_c=\frac{-\alpha_1-\sqrt{\alpha_1^2-2\alpha_2E_g}}{\alpha_2}.
\end{equation}
The relationship given in Eq~(\ref{eq:Pc}) implies that
  the value of $P_c$ could be obtained
  from the equilibrium band gap $E_g$ (for which the experimental value is available)
   and the band-gap pressure coefficients $\alpha_1$ and $\alpha_2$ (for which the experimental values are lacking).
Thus, experimental determination of $\alpha_1$ and $\alpha_2$
  is called for.

%
%
\begin{figure}
  \begin{center}
    \resizebox{0.5\textwidth}{!}{
      \includegraphics{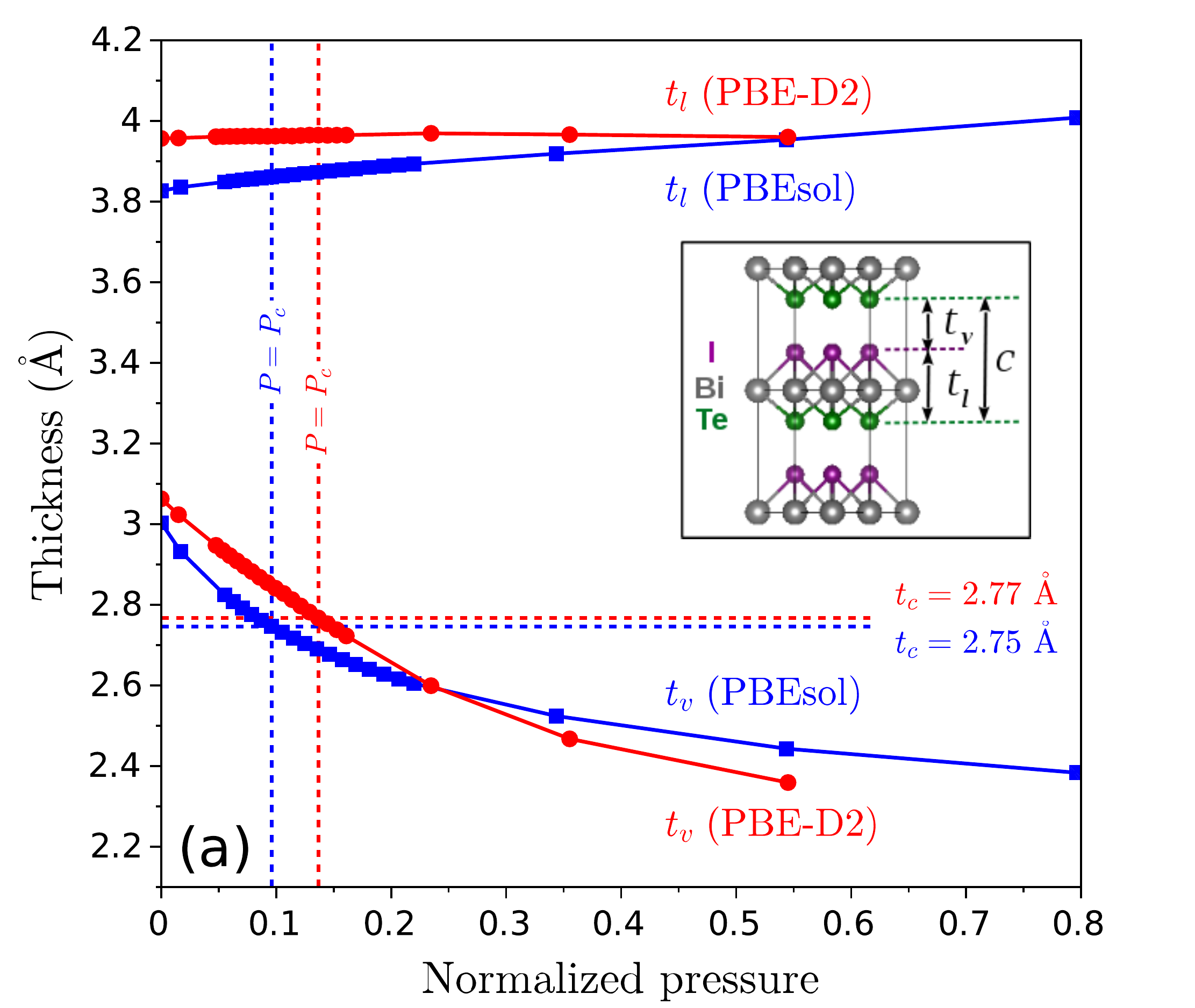}
    }
    \resizebox{0.5\textwidth}{!}{
      \includegraphics{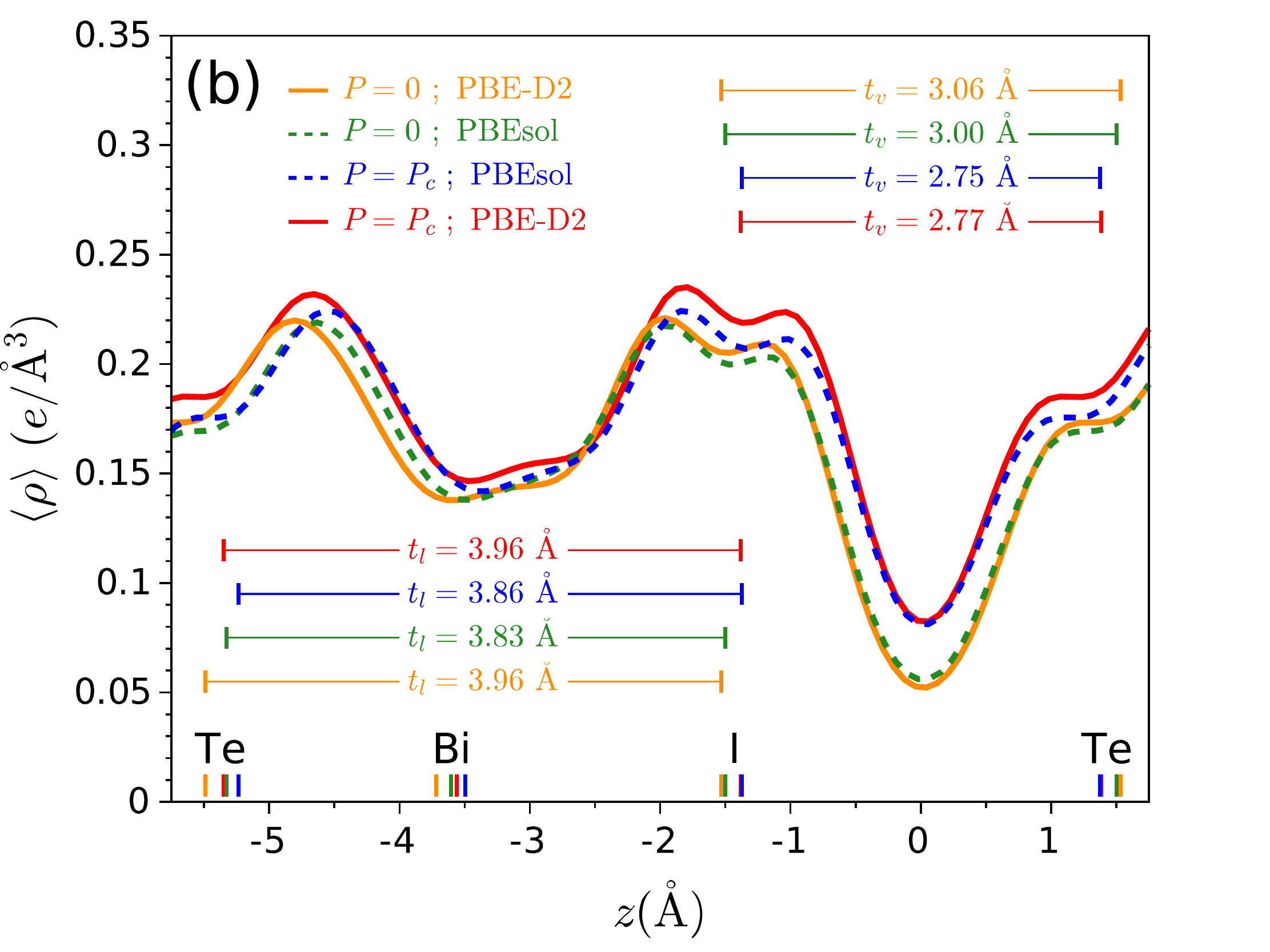}
    }
    \resizebox{0.5\textwidth}{!}{
      \includegraphics{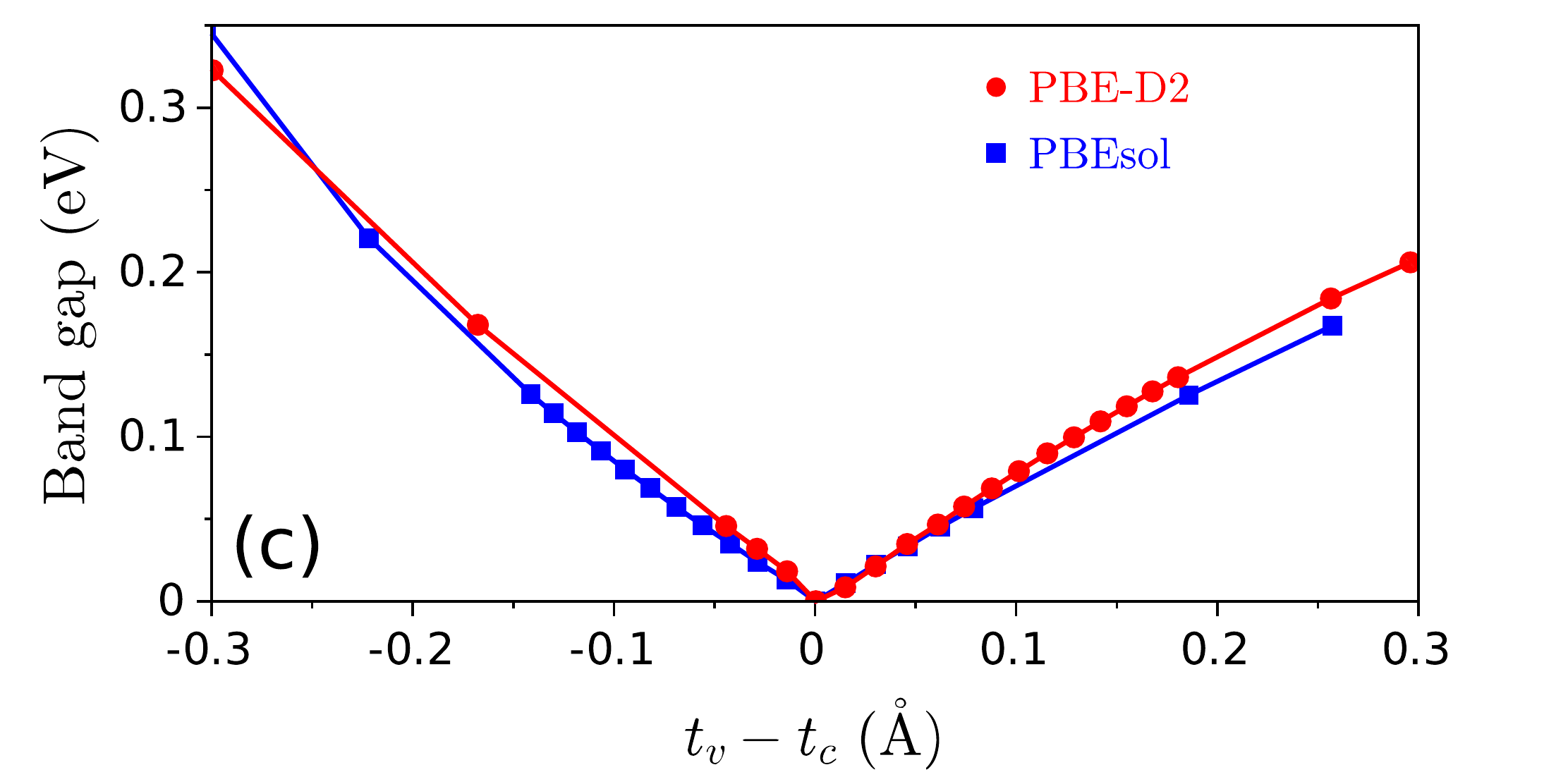}
    }
  \end{center}
  \vspace*{-0.5cm}
  \caption{(a) The layer thickness $t_l$ and the width $t_v$ of the van der Waals gap,
               which are depicted in the inset,
               are plotted with respect to the normalized pressure $P/K_0$.
               The vertical and horizontal dashed lines mark the values of $P_c/K_0$ and $t_c$, respectively.
           (b) The $ab$ planar average $\langle \rho \rangle$ of the charge density is plotted
               as a function of the position $z$ along the $c$-axis
               for zero pressure [the orange solid (PBE-D2) and green dashed (PBEsol) curves]
               and for $P=P_c$   [the    red solid (PBE-D2) and  blue dashed (PBEsol) curves].
           (c) The band-gap energy versus the difference $t_v-t_c$.
  }
  \label{f:tv}
\end{figure}

It was proposed that
  an observed minimum in $c/a$ in the pressure range of 2.0--2.9~GPa
  is an indicator of the topological phase transition.\cite{xi13,chen13}
It is thus interesting to explore how the lattice parameter ratio $c/a$ varies with pressure.
The plots of $c/a$ versus the normalized pressure are therefore given in Fig.~\ref{f:eos}(c)
  where the experimental values (the green diamonds) are seemingly {\it bracketed} by the blue and red curves
  passing through the PBEsol-calculated and dispersion-corrected (PBE-D2) points, respectively.
As a matter of fact
  the results of PBEsol and PBE-D2 calculations are in better agreement with the experimental data
  in regard to the pressure variation of the lattice parameters $c$ and $a$, respectively,
  cf. Fig.~S2 (Ref.~\onlinecite{supmat}).
It should be remarked that
  the PBEsol-calculated points (the blue squares)
  form a curve that lies virtually parallel to the experimental points,
  and accordingly possess a minimum.
On contrary, the curve passing through the dispersion-corrected (PBE-D2) points (the red circles)
  shows no minimum.
It is thus clear that the results of the PBEsol (PBE-D2) calculations
  are {\it in line (conflict) with} the experimental trend
  in regard to the pressure variation of the lattice parameter ratio $c/a$.
Despite this disagreement between the PBEsol and PBE-D2 results,
  the closure of the band gap at the critical pressure
  occurs within {\it not only} the PBEsol {\it but also} PBE-D2 description,
  which reveals that the pressure variation of the band gap is {\it not} in correlation with that of $c/a$.
A comparative inspection of Figs.~\ref{f:eos}(b) and \ref{f:eos}(c)
  indeed shows that the transition to the zero-gap state
  does {\it not} correspond to the minimum of $c/a$.

\begin{figure*}
  \begin{center}
    \resizebox{0.84\textwidth}{!}{
      \includegraphics{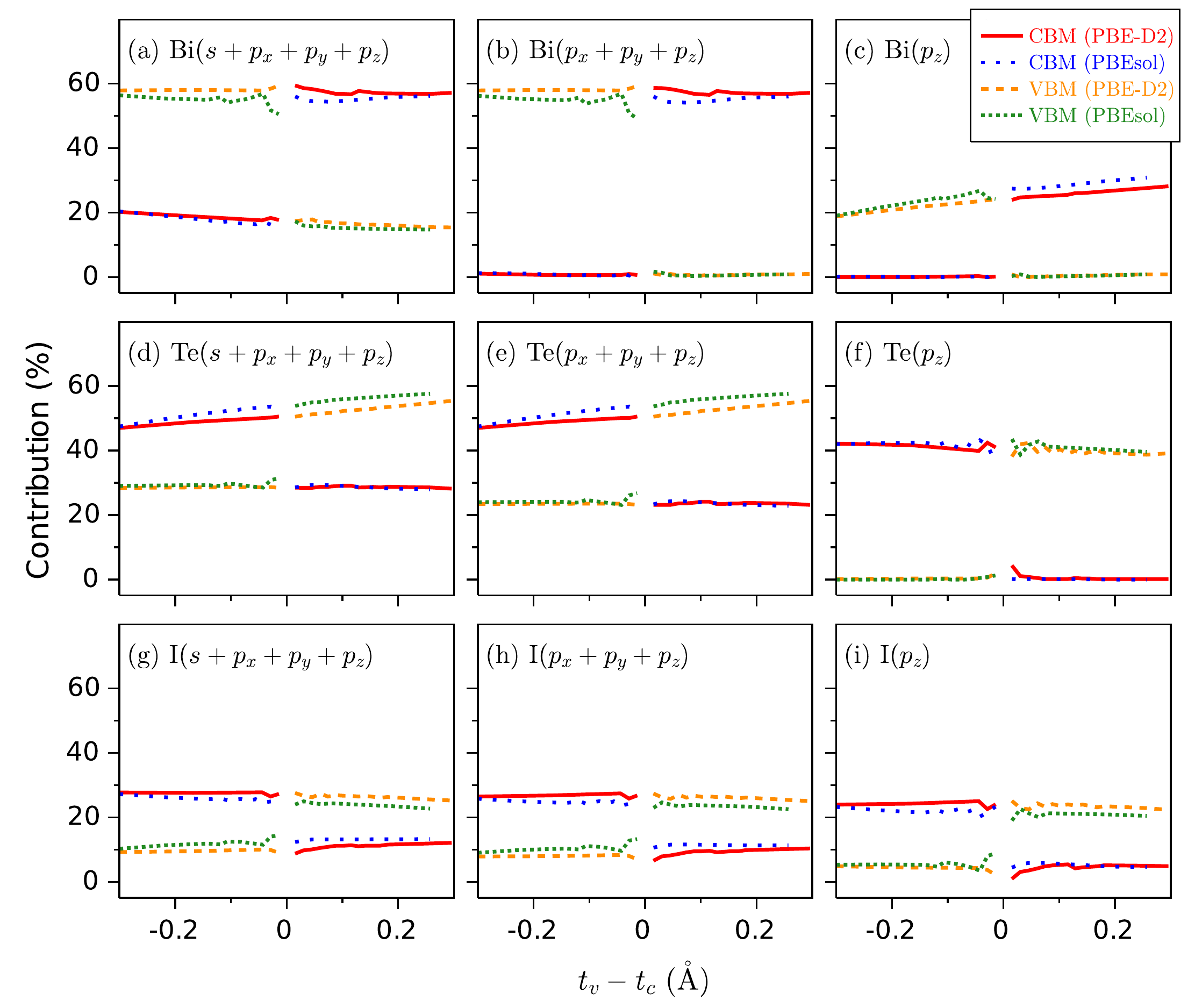}
    }
  \end{center}
  \vspace*{-0.5cm}
  \caption{ The variation of the $s+p_x+p_y+p_z$ [(a), (d) and (g)], 
                                                 $p_x+p_y+p_z$ [(b), (e) and (h)] and 
                                                         $p_z$ [(c), (f) and (i)] contributions
                                                       from Bi [(a)-(c)],
                                                            Te [(d)-(f)] and
                                                            I  [(g)-(i)] atoms
           to the VBM and CBM wave functions
           with the difference $t_v-t_c$.
           Note that the latter is positive (negative) for pressures lower (higher) than $P_c$, which is {\it zero} for $P=P_c$.
          }
  \label{f:contri}
\end{figure*}

We find that
  the width of the vdW gap,
  viz. the thickness $t_v$ of the vacuum region,
  rather than the lattice parameter ratio $c/a$, 
  serves as an adequate structural parameter
  for studying the variation of the BiTeI band-gap energy with pressure,
  as will be explained now.
Figure~\ref{f:tv}(a) shows the variation of $t_v$
  and the layer thickness $t_l$ with the normalized pressure,
  where the inset displays the division of the lattice parameter $c$ into $t_v$ and $t_l$.
While the pressure increases,
  $t_l({\rm PBE\mbox{-}D2})$ remains virtually constant
  whereas $t_l({\rm PBEsol})$ tends to increase slowly. 
The slight increase of the layer thickness as a result of compression
  was observed in TiS$_2$, viz. a layered material with a vdW gap,
  which is associated with an increase of the electronic charge in the interlayer region,
  indicating enhanced {\it metallic} rather than van der Waals bonding between the layers.\cite{allan98}
In accordance with this,
  we think that
  the qualitative description of the pressure variation of $t_l$
  within the PBEsol, rather than PBE-D2, approach
  is {\it realistic} (in the low-pressure regime).
The latter is also supported by the fact that
  the results of PBEsol, rather than PBE-D2, calculations are in excellent agreement with 
  the measured pressure variation of the lattice parameter $c$,
  cf. Fig.~S2 (Ref.~\onlinecite{supmat}).
As long as $c=t_l+t_v$,
  the effect of pressure on $c$ is manifested primarily (solely) as the compression of the vdW gap
  according to the PBEsol (PBE-D2) calculations.
While $t_v$ is getting reduced,
  the electronic charge density $\rho(\mathbf r)$ in the vdW gap region increases,
  as illustrated in Fig.~\ref{f:tv}(b)
  where the $ab$ planar average $\langle \rho \rangle$ of the charge density
  is plotted for zero pressure and for $P=P_c$.
This means that
  the interlayer metallic bonding
  becomes stronger as $t_v$ decreases.
The decrease in $t_v({\rm PBEsol})$, compared to that in $t_v({\rm PBE\mbox{-}D2})$,
  is more pronounced in the low-pressure regime,
  as seen in Fig.~\ref{f:tv}(a).
Accordingly,
  the transition to the zero-gap state occurs at a relatively lower (higher) pressure
  within the PBEsol (PBE-D2) approach.
This explains why $P_c^{\rm PBEsol}$ is smaller than $P_c^{\rm PBE\mbox{-}D2}$.

The band gap diminishes
  when the width $t_v$ of the vdW gap is reduced to a critical value $t_c$,
  which is marked by the horizontal dashed lines in Fig.~\ref{f:tv}(a).
It is remarkable that
  $t_c$
  takes practically the same value
  in {\it both} PBEsol and PBE-D2 approaches,
  i.e., $t_c\approx 2.76$ \AA.
Hence the transition to the zero-gap state occurs
  at the {\it same} critical thickness $t_c$
  according to the {\it two} approximations (PBEsol and PBE-D2) employed in this study.
It is also striking that
  the PBEsol and PBE-D2 calculations
  yield almost {\it identical} $\langle \rho \rangle$ in the vdW gap region
  for zero pressure as well as for $P=P_c$,
  which is discernible in Fig.~\ref{f:tv}(b);
  compare
  the orange solid  ($P=0$;   PBE-D2) curve to
  the  green dashed ($P=0$;   PBEsol) curve, and
  the    red solid  ($P=P_c$; PBE-D2) curve to
  the   blue dashed ($P=P_c$; PBEsol) curve.
Thus the PBEsol and PBE-D2 approaches
  that are in conflict to some degree (as pointed out above)
  yield {\it consistent} predictions
  concerning the electronic structure and the interlayer metallic bonding in BiTeI under pressure.
This is further evidenced in Fig.~\ref{f:tv}(c)
  where the band-gap energy is plotted with respect to the difference $t_v-t_c$.
It is seen that
  the calculated (PBEsol and PBE-D2) values of the band gap
  follow not only {\it qualitatively} but also {\it quantitatively} the same trend.
Thus the difference between the predictions of the PBEsol and PBE-D2 approaches
  does {\it not} pertain to the variation of the electronic structure of BiTeI under pressure,
  which actually pertains only to the overestimation (PBEsol) and underestimation (PBE-D2) of the compressibility of BiTeI.

The predictions of the PBEsol and PBE-D2 approaches
  are further compared to each other 
  in the matter of the inversion of the character of the VBM and CBM states,
  which was mentioned in the beginning of this article, cf. Refs.~\onlinecite{bahramy12, xi13, ideue14, park15}.
The angular-momentum-resolved ($s$, $p_x$, $p_y$, $p_z$) contributions
  from Bi, Te, and I atoms
  are computed
  by projecting the VBM and CBM wave functions onto spherical harmonics
  within a sphere around each atom. 
Figure~\ref{f:contri} shows plots of
  the $s+p_x+p_y+p_z$ (the left   panels), 
        $p_x+p_y+p_z$ (the center panels) and 
                $p_z$ (the right  panels) contributions
  from Bi (the  upper panels),
       Te (the middle panels) and
       I  (the  lower panels) atoms
           to the VBM and CBM wave functions
           as a function of $t_v-t_c$.
It is seen in Figs.~\ref{f:contri}(g)-\ref{f:contri}(i) that
  the total,
      total $p$ and
            $p_z$ contributions to the character of either VBM or CBM wave function
  from the I atom
  remain roughly constant regardless of the value of $t_v-t_c$, viz. the degree of compression.
The total $p$ contribution from the Bi atom
    is also roughly constant, as seen in Fig.~\ref{f:contri}(b).
On the other hand,
   Figs.~\ref{f:contri}(a), \ref{f:contri}(c)-\ref{f:contri}(f)
   show some variation.
For example, 
  Fig.~\ref{f:contri}(a) [\ref{f:contri}(d)] shows that
  the overall contribution from Bi [Te] atoms
  is slightly increasing [decreasing]
  as the degree of compression increases.
The contribution from Bi-$p_z$ [Te-$p_z$]
  decreases [increases]
  at the expense of increasing contributions from Bi-$p_x$ and Bi-$p_y$ [Te-$p_x$ and Te-$p_y$],
  cf. Figs.~\ref{f:contri}(a) and \ref{f:contri}(b) [Figs.~\ref{f:contri}(e) and \ref{f:contri}(f)],
  while $t_v$ is getting reduced.
Accordingly,
  the character of the CBM (VBM) wave function is mostly of Bi-$p$ (Te-$p_z$+I-$p_z$) for $t_v-t_c > 0 $,
  which is of Te-$p_z$+I-$p_z$ (Bi-$p$) for $t_v-t_c < 0 $.
This demonstrates clearly the inversion of the VBM and CBM states that occurs at $t_v-t_c=0$.
As a matter of fact
  the inversion of the character of the VBM and CBM wave functions
  is easily noticed in {\it any} panel of Fig.~\ref{f:contri},
  thanks to the discontinuity of the plotted curves at $t_v-t_c=0$.
It is also readily noticeable that
  the PBEsol-calculated and dispersion-corrected (PBE-D2) curves 
  follow {\it qualitatively} and {\it quantitatively} (within a few percent) the same trend
  in {\it all} panels of Fig.~\ref{f:contri}.
Hence the results of the PBEsol and PBE-D2 calculations are in agreement
  as regards
  the inversion of the character of the VBM and CBM wave functions.

In conclusion, the evolution of the electronic structure of BiTeI under compression
  is studied by employing semilocal (PBEsol) and dispersion-corrected (PBE-D2) density-functional calculations.
A comparative investigation of the results of these calculations
  confirms that
  the band-gap energy of BiTeI decreases till it attains a minimum value of {\it zero}
  at a critical pressure, after which it increases again,
  which was reported in a number of recent studies.\cite{bahramy12, xi13, ideue14, park15}
The critical pressure is found to be lower than
  the pressure at which BiTeI undergoes a structural phase transition,
  implying that the closure of the band gap would not be hindered by a structural transformation.
In addition, the band-gap pressure coefficients of BiTeI are computed, and
  an expression of the critical pressure is devised in terms of these coefficients, cf. Eq.~(\ref{eq:Pc}).
It is to be emphasized that the latter enables one to estimate 
  the critical pressure from the equilibrium band gap and its pressure coefficients.
It is exposed that
  the essential difference between the results of PBEsol and PBE-D2 calculations
  pertains to the prediction of the compressibility of BiTeI.
Nevertheless,
  the effect of pressure on the atomic structure of BiTeI is found to be manifested primarily
  as the reduction of the width of the van der Waals gap
  according to both types of calculations.
It is further revealed that
  the PBEsol and PBE-D2 approaches
  yield consistent predictions concerning
  the variation of band-gap energy with respect to the width of the van der Waals gap.
Consequently, it is shown that
  the calculated (PBEsol and PBE-D2) band-gap energies follow {\it qualitatively} and {\it quantitatively} the same trend
  within the {\it two} approximations employed here, and
  the transition to the zero-gap state occurs at the {\it same} critical width of the van der Waals gap.

\bigskip
We thank Xiaoxiang Xi for providing the experimental data used in Figs.~\ref{f:eos}(a), \ref{f:eos}(c), and S2.
The numerical calculations reported here were carried out at the High Performance and Grid Computing Center (TRUBA Resources) of TUBITAK ULAKBIM.

\appendix*

\renewcommand{\thefigure}{A\arabic{figure}}
\renewcommand{\thetable}{A}

\setcounter{figure}{0}
\setcounter{table}{0}

\section{Crystal structure calculations using the optB86b-vdW functional}

\begin{figure}
  \begin{center}
    \resizebox{0.48\textwidth}{!}{%
      \includegraphics{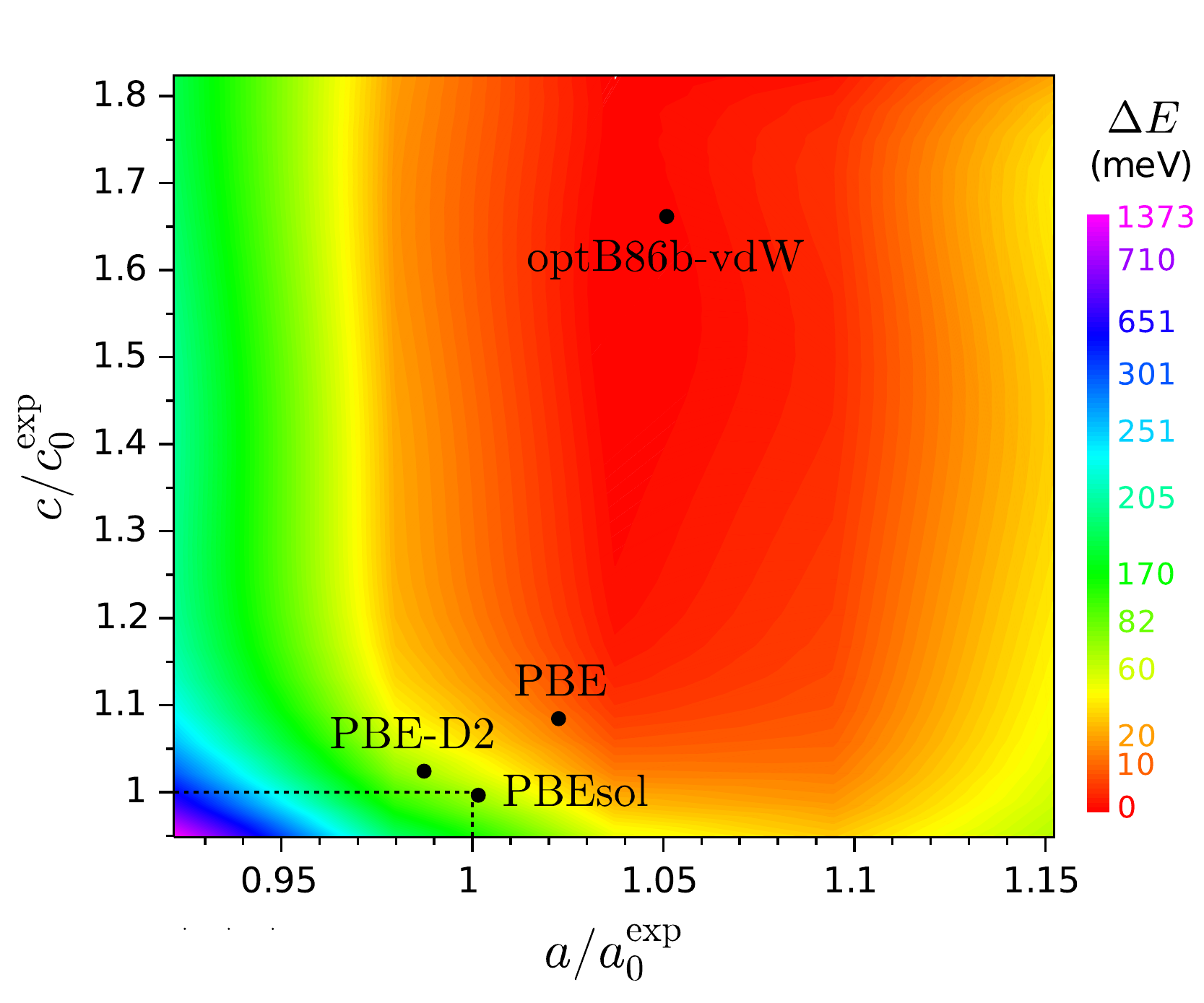}
    }
  \end{center}
  \vspace*{-0.5cm}
  \caption{
          Colored contour plot of the energy difference $\Delta E$
            computed using the optB86b-vdW functional
            as a function of $a/a_0^{\rm exp}$ and $c/c_0^{\rm exp}$.
          The solid black circles mark the values of the ratios $a_0/a_0^{\rm exp}$ and $c_0/c_0^{\rm exp}$
            obtained via various approximations, i.e., PBE, PBE-D2, PBEsol, and optB86b-vdW.
          }
   \label{f:optB86b}
\end{figure}

We computed the total energy $E$
  as a function of the lattice parameters $a$ and $c$
  using the optB86b-vdW functional,\cite{klimes11}
  and determined the energy minimum $E_{\rm min}$.
A colored contour plot of the energy difference $\Delta E=E-E_{\rm min}$
  versus $a/a_0^{\rm exp}$ and $c/c_0^{\rm exp}$ 
  is given in Fig.~\ref{f:optB86b},
  where $a_0^{\rm exp}$ and $c_0^{\rm exp}$ denote the experimental values of
  the equilibrium lattice parameters of $a_0$ and $c_0$, respectively.
There appears to exist an anomalously flat energy plateau
  that is the red portion of the plot in Fig.~\ref{f:optB86b}.
Within this plateau,
  the optB86b-vdW functional
  results in a great overestimation in $c_0$ ($a_0$) by 66~\% (5~\%).
In other words, crystal structure calculations using the optB86b-vdW functional yield
  an {\it unacceptably unrealistic (unexpectedly inaccurate)} prediction for $c_0$ ($a_0$),
  cf. the solid black circles in Fig.~\ref{f:optB86b},
  which is clearly {\it not} the case with the other calculation (PBE, PBE-D2, PBEsol) results.
Owing to this difficulty,
  the resolution of which is beyond the scope of this paper,
  we avoid employing the optB86b-vdW functional in electronic structure calculations.

\clearpage

\section*{Supplemental Material}

\setcounter{page}{1}
\setcounter{figure}{0}
\setcounter{table}{0}
\renewcommand{\thepage}{S-\arabic{page}}
\renewcommand{\thefigure}{S\arabic{figure}}
\renewcommand{\thetable}{S\arabic{table}}

\begin{itemize}
\item Figure~S1 displays the fourth- and fifth-order Birch-Murnaghan fits for the PBEsol-calculated and dispersion-corrected (PBE-D2) energy-volume curves, respectively.
\item Figure~S2 shows the plots of $a/a_0$ and $c/c_0$ versus the normalized pressure $P/K_0$,
          where $a_0$ and $c_0$ denote the equilibrium values of the lattice parameters $a$ and $c$, respectively.
\end{itemize}

\clearpage

\begin{figure*}
  \begin{center}
    \resizebox{0.75\textwidth}{!}{
      \includegraphics{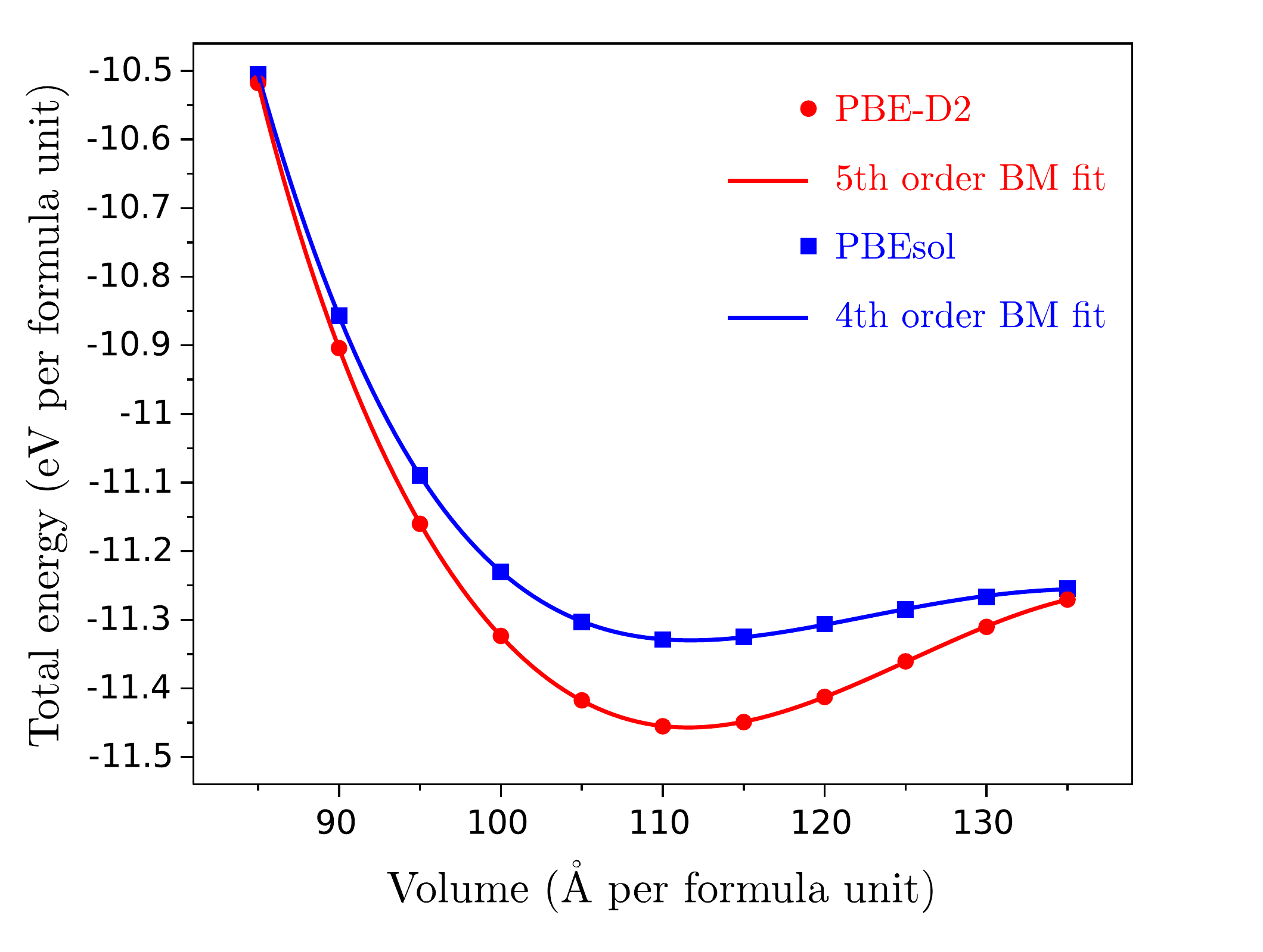}
    }
  \end{center}
  \vspace*{-0.5cm}
  \caption{ The energy-volume curves obtained via the PBEsol and PBE-D2 calculations.
           It is discernible that the PBEsol-calculated and dispersion-corrected (PBE-D2) curves are reproduced well by fourth- and fifth-order Birch-Murnaghan (BM) fits, respectively.
          }
\end{figure*}

\begin{figure*}
  \begin{center}
    \resizebox{0.88\textwidth}{!}{
      \includegraphics{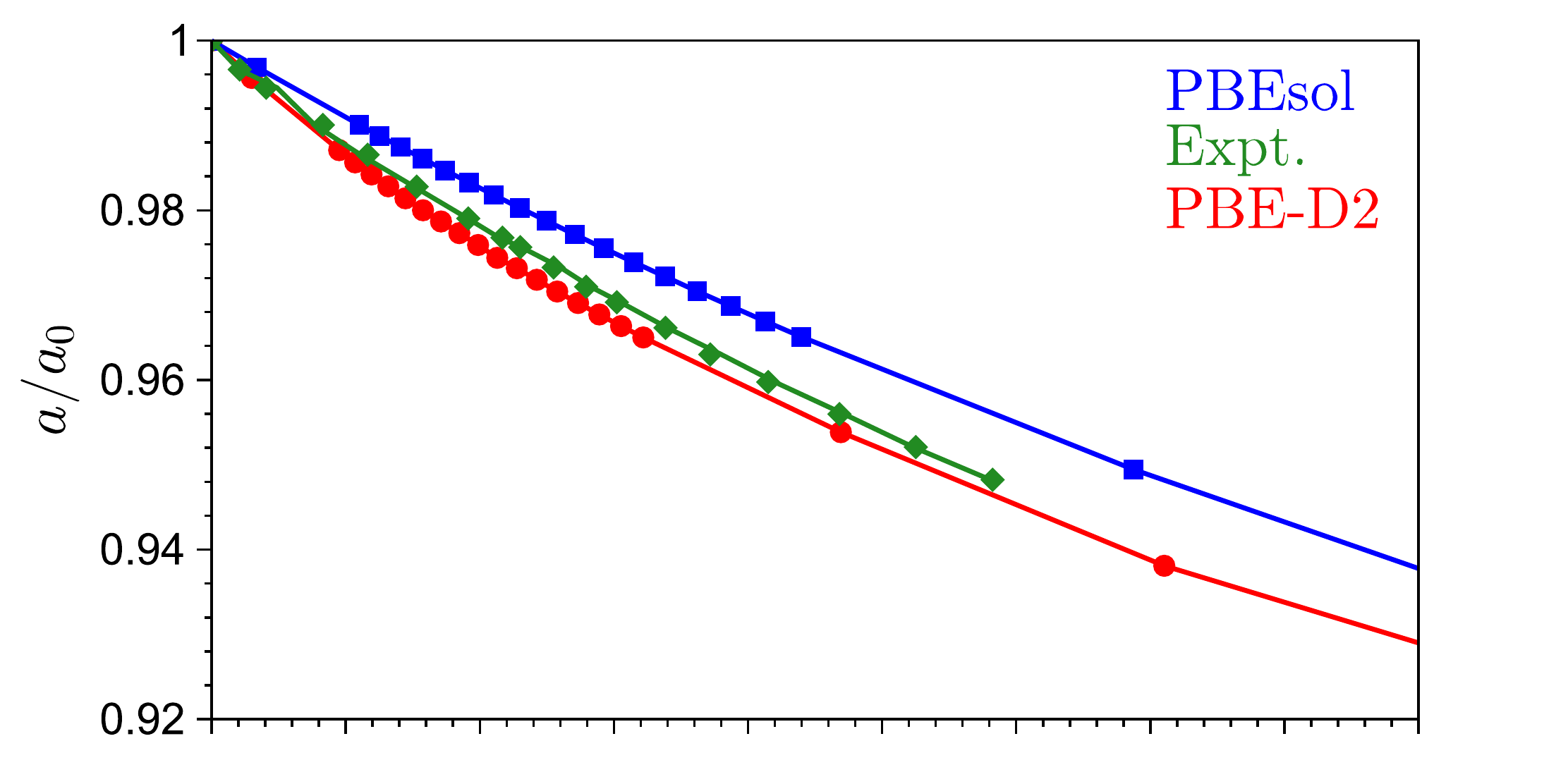}
    }
    \resizebox{0.88\textwidth}{!}{
      \includegraphics{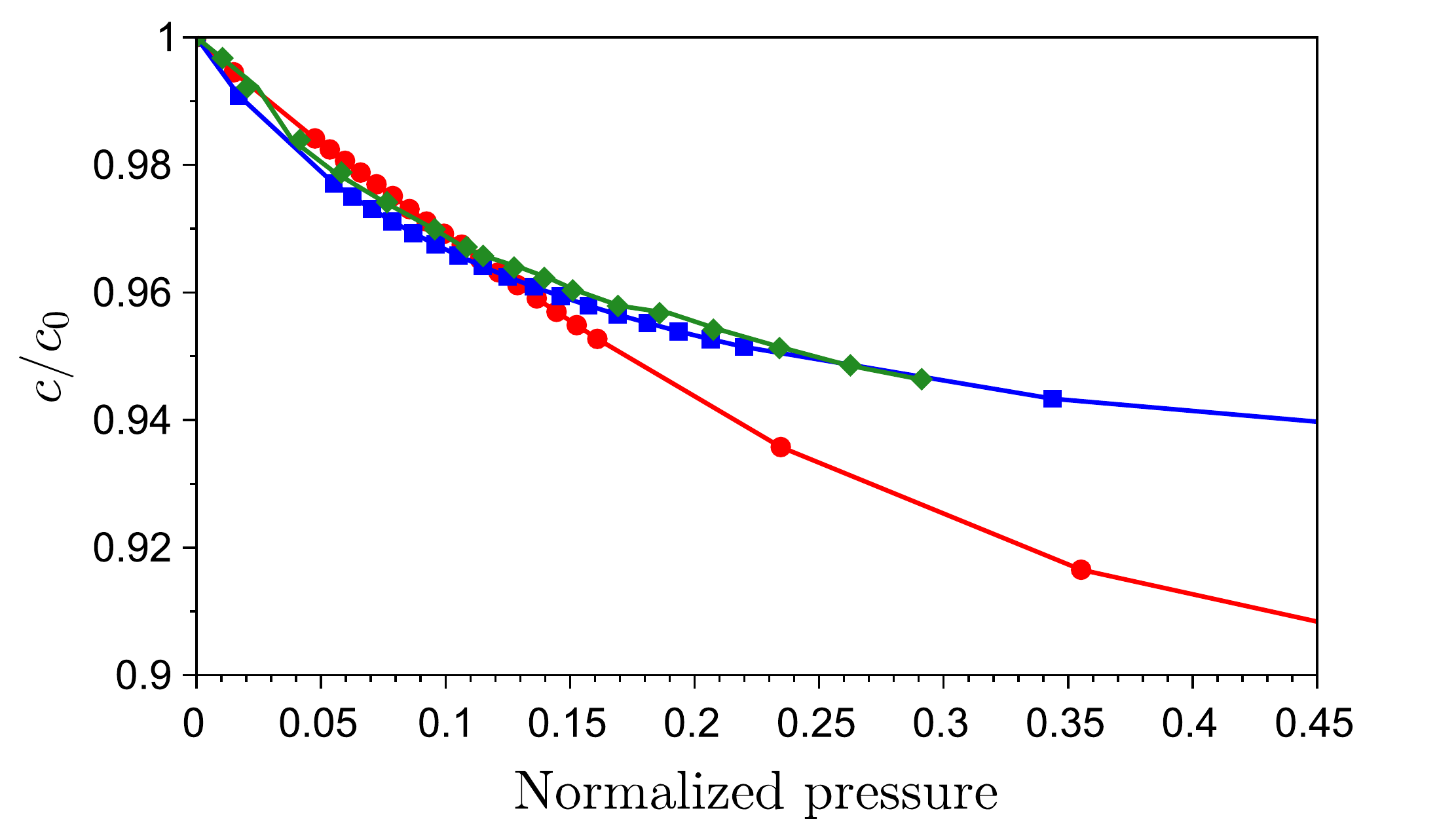}
    }
  \end{center}
  \vspace*{-0.5cm}
  \caption{ The plots of $a/a_0$ (top panel) and $c/c_0$ (bottom panel) versus the normalized pressure $P/K_0$,
           where $a_0$ and $c_0$ denote the equilibrium values of the lattice parameters $a$ and $c$, respectively.}
\end{figure*}

\end{document}